\documentclass[12pt]{iopart}

\usepackage{times}

\usepackage{graphicx}

\def\leqsim{\mathbin{\;\raise1pt\hbox{$<$}\kern-8pt\lower3pt\hbox{\small$\sim$}\;}}
\def\geqsim{\mathbin{\;\raise1pt\hbox{$>$}\kern-8pt\lower3pt\hbox{\small$\sim$}\;}}

\def\MXN#1{\mbox{$ m_{\tilde{\chi}^0_#1}                                $}}

\def\MXC#1{\mbox{$ m_{\tilde{\chi}^{\pm}_#1}                            $}}

\def\XPM#1{\mbox{$ \tilde{\chi}^{\pm}_#1                                $}}
\def\XN#1{\mbox{$ \tilde{\chi}^0_#1                                     $}}

\def\p#1{\mbox{$ \mbox{\bf p}_1                                         $}}

\newcommand{\tanb}    {\mbox{$ \tan \beta                                  $}}

\newcommand{\msmur}   {\mbox{$ m_{\tilde{\mu}_R}                           $}}

\newcommand{\stau}     {\mbox{$ \tilde{\tau}                               $}}
\newcommand{\stauo}     {\mbox{$ \tilde{\tau}_1                            $}}

\newcommand{\mstau}   {\mbox{$ m_{\tilde{\tau}}                            $}}
\newcommand{\mstauo}   {\mbox{$ m_{\tilde{\tau}_1}                         $}}

\newcommand{\snu}     {\mbox{$ \tilde\nu                                   $}}
\newcommand{\msnu}    {\mbox{$ m_{\tilde\nu}                               $}}
\newcommand{\msell}   {\mbox{$ m_{\tilde{\mathrm e}_L}                     $}}
\newcommand{\mselr}   {\mbox{$ m_{\tilde{\mathrm e}_R}                     $}}
\newcommand{\sell}   {\mbox{$ {\tilde{\mathrm e}_L}                     $}}
\newcommand{\selr}   {\mbox{$ {\tilde{\mathrm e}_R}                     $}}

\newcommand{\sle}     {\mbox{$ \tilde{\ell}                                $}}

\newcommand{\stq}     {\mbox{$ \tilde {\mathrm t}                          $}}
\newcommand{\mstq}    {\mbox{$ m_{\tilde {\mathrm t}}                      $}}
\newcommand{\sbq}     {\mbox{$ \tilde {\mathrm b}                          $}}
\newcommand{\msbq}    {\mbox{$ m_{\tilde {\mathrm b}}                      $}}

\newcommand{\hn}      {\mbox{$ {\mathrm h}^0                               $}}
\newcommand{\Zn}      {\mbox{$ {\mathrm Z}                                 $}}

\newcommand{\ee}      {\mbox{$ {\mathrm e}^+ {\mathrm e}^-                 $}}

\newcommand{\GeV}     {\mbox{$ {\mathrm{GeV}}                              $}}

\newcommand{\GeVcc}   {\mbox{$ {\mathrm{GeV}}/c^2                          $}}

\newcommand{\TeVcc}   {\mbox{$ {\mathrm{TeV}}/c^2                          $}}
\newcommand{\pbi}     {\mbox{$ {\mathrm{pb}}^{-1}                          $}}
\newcommand{\MZ}      {\mbox{$ m_{\mathrm Z}                               $}}

\def\PR#1#2#3{{\rm Phys.~Rep.} {\bf#1} (19#2) #3}

\def\CPC#1#2#3{{\rm Comp.~Phys.~Comm.} {\bf#1} (19#2) #3}

\def    \DM          {\mbox{$\Delta M$}}
\def    \missEt      {\ifmmode{/\mkern-11mu E_t}\else{${/\mkern-11mu E_t}$}\fi}
\def    \missE       {\ifmmode{/\mkern-11mu E}\else{${/\mkern-11mu E}$}\fi}
\def    \missp       {\ifmmode{/\mkern-11mu p}\else{${/\mkern-11mu p}$}\fi}
\def    \misspt      {\ifmmode{/\mkern-11mu p_t}\else{${/\mkern-11mu p_t}$}\fi}

\begin{document}

\title{Understanding   SUSY limits from LEP} 

\author{Anna Lipniacka}

\address{University of Stockholm, Fysikum, Alba Nova Stockholm
Center for Physics, Astronomy and Biotechnology,
S - 106 91 Stockholm, Sweden}

\begin{abstract}
LEP results  have constrained  heavily the
Minimal  Supersymmetric Standard  Model, while providing
hints for light Higgs boson and for ``SUSY-assisted'' gauge
couling unification.    In this  paper the  results
obtained  at  LEP  within  two
scenarios, the gravity-mediated  MSSM  framework and  the
minimal  SUGRA scenario are presented. 
Model-dependence and coverage  of LEP results is discussed.

\article{Preprint USIP-2/2002}{}

\end{abstract}



\section{Introduction}
\label{sec:intro}

Supersymmetry (SUSY) is believed to be one of the most attractive scenarios
for physics beyond the Standard Model. In the last few years around 150 papers
on experimental searches for SUSY were
 published, out of which around 100 were related to the LEP
results. This large number of papers
reflects perhaps as well the large number of free parameters relevant to
SUSY models at the presently explored energy scale.  
LEP is well suited to explore corners of 
SUSY models in a relatively assumption independent way.

        In this paper the results obtained by LEP experiments  within 
the gravity-mediated 
constrained MSSM framework and  the minimal SUGRA scenario
are presented, emphasis is put on the model dependence of the
exclusion. See \cite{michael} for a recent  general review. 
 
         In the  Minimal Supersymmetric extension of the Standard Model (MSSM) 
\cite{sugpeda},
each Standard Model particle has a supersymmetric partner with the same
couplings and with spin differing by 
$\hbar$/2.  Large corrections
to the Higgs mass from interactions involving  
virtual particles (heavy quarks in particular) are partially cancelled
due  to their superpartners. If they are lighter
than  1-10 \TeVcc\  this solves the so called
hierarchy problem \cite{SUSY2}. Moreover, supersymmetric particles
modify the  energy  dependence of the electromagnetic, weak and strong
coupling constants, and help them to unify at the scale of  
around $10^{15}$~\GeV  \cite{wim}.

          The Higgs sector of the MSSM has to be extended to two complex
Higgs doublets $H_1,H_2$ responsible for  giving masses 
to the up and down-type fermions. 
Five physical Higgs boson mass states remain after the
Electroweak Symmetry breaking.
The lightest scalar neutral Higgs boson $\hn$ and the heavier pseudoscalar
neutral Higgs boson   $A$ are of interest for this paper.  
On the tree level, masses of the Higgs bosons depend on 
just two  parameters,
which can be chosen as  tan$\beta$, the ratio of vacuum expectation 
values of the two Higgs doublets,
and $m_A$. In particular   $m_h < m_Z*|cos2\beta| $ \footnote{For 
$m_A>>m_Z$, $m_{\hn(tree)} \sim  m_Z*|cos2\beta|/(1+m_Z^2/m_A^2) $, 
and for \tanb $\geqsim$ 10, $m_{\hn (tree)} \sim m_Z*|cos2\beta| $},  
however due to radiative
corrections mentioned above (which depend on the top quark mass, and 
on the mass terms of the superpartners of heavy quarks),
the upper limit on the mass of  the lightest Higgs boson 
grows to  $m_h \leqsim$~135 \GeVcc\ \cite{higtherlim,carwig}.\\
     If $m_A \geqsim$ 150~\GeVcc\ the lightest supersymmetric Higgs boson
resembles very much the one of the Standard Model.
Precise  electroweak
measurements  \cite{lepew} 
suggest that the Higgs boson is relatively light
\footnote{The central value moves to $\sim$ 110~\GeVcc\ if
the top quark mass is assumed to be one standard
deviation above its central value}, $m_h = 88^{+53}_{-36}$\GeVcc, 
well in the range of the MSSM prediction. 
Searches for the Standard Model
like Higgs boson at LEP \cite{higdelphi,higlep} 
set a lower limit for $m_h$, 
$m_h>$114.4 \GeVcc\ ( if \tanb$<$ 6, or  $m_A>$120~\GeVcc),
constraining heavily the MSSM. The 1.7$\sigma$ ``excess'' observed at LEP 
\cite{fabiola} of events compatible with  
production of the Standard Model Higgs boson with   
$m_h \sim 114-117 $ \GeVcc,  together with the EW constraints, makes
low $m_h$, just above the reach of LEP, quite probable.

     The MSSM provides a phenomenologically interesting wealth of superpartners
of the Standard Model particles.
Supersymmetric partners of gauge 
and Higgs bosons (gauginos and higgsinos) 
mix to 
realize  four neutral mass states, neutralinos, $\XN{i}_{: i=1,4}$,
and four charged mass states, charginos, \XPM{1},\XPM{2}.
Superpartners of left-handed and right-handed fermions,
``right-handed'' and ``left-handed'' scalar quarks (squarks) and 
scalar leptons (sleptons) can mix. This leads to the off-diagonal
``left-right'' terms in their mass matrices and induces an 
additional  mass splitting
between the lighter and the heavier state.

        While the Higgs sector is well constrained in the MSSM, very little
can be said about the superpartners mass spectrum unless one makes
some additional assumptions.
As no superpartners were found so far
the Supersymmetry has   to be
broken. The pattern  of the  sparticle  mass spectrum 
depends primarily on
the mechanism of its  breaking.
 
       In the models with  gravity mediated supersymmetry breaking which
will be discussed in this paper,  the lightest neutralino (\XN{1}) is  
usually the
Lightest Supersymmetric Particle (LSP). If R-parity 
\footnote{$R$-parity is a multiplicative quantum number defined as 
$R=(-1)^{3(B-L)+2S}$
where $B$, $L$ and $S$ are the baryon number, the lepton number and the spin
of the particle, respectively. SM particles have R=+1 while 
their SUSY partners have $R=-1$}
is conserved the  LSP does not decay, and it is an ideal cold dark 
matter candidate
\cite{larsb}.Constraints on models with broken R-parity were discussed in
\cite{r-parity} and are thus not discussed in this paper.
 
       Experimental searches motivated by the  MSSM with R-parity conservation 
and  gravity-mediated
supersymmetry breaking  exploit features of the model independent of 
further assumptions, like the 
strength of superpartner  couplings to  the gauge bosons, pair-production of
sparticles, and the missing energy and momentum signature due to escaping LSPs 
in the final state.

         However, to cover ``pathological'' situations with   final 
states which cannot be efficiently
detected  or situations  where the  production cross-sections are low, 
or finally to achieve more predictivity
and set limits on masses of the sparticles which are not directly 
observable (e.g. the LSP in the R-parity conserving model), 
additional model assumptions have to be made. In this paper 
two ``flavours'' of such
constraining assumptions are discussed (see section
\ref{sec:models}): the constrained MSSM with non-universal Higgs 
parameters (CMSSM with nUHP),
which is often used to interpret LEP results, and an even more constrained
minimal SUGRA scenario (mSUGRA) \footnote{ The definition of mSUGRA used
in this paper corresponds to what is called CMSSM with universal Higgs
masses in \protect{\cite{ellislight,benchellis,roszk}}}, 
often used to interpret 
Tevatron results  and
for  benchmark searches at  future colliders \cite{benchellis}.
It is shown in  section \ref{sec:limits}  that in both models  LEP results
can be used to exclude sparticles much beyond the kinematic limit of LEP.\\
   Perspectives to find sparticles at the Tevatrons Run II in view of 
limits from LEP are discussed in \cite{mytevpaper}.

\section{ The models: CMSSM with nUHP and mSUGRA}
\label{sec:models}

   To make the MSSM 
more predictive,  
the unification of some parameters at a high mass scale typical 
of Grand Unified Theories (GUT) can be assumed. 
In this section, approximate relations between the model parameters and
the superparners masses  which are important to 
understand the experimental limits
will be quoted without explanations. For a more complete 
information see  e.g. \cite{sugpeda}.

\noindent
\subsection{CMSSM with nUHP}
\label{sec:cmssm}

As well as  the already mentioned \tanb\ and $m_A$, the following parameters
are relevant in the constrained  MSSM with non-universal Higgs parameters:

\begin{itemize}

\item
\mbox{\boldmath$\mu$}, the Higgs mass parameter,

\item
\mbox{\boldmath $M_1,M_2,M_3$}, the  $U(1)\times SU(2)\times SU(3)$ 
gaugino masses 
at the electroweak (EW) scale.
Gaugino mass unification at the GUT scale is assumed, with a common gaugino 
mass
of $\boldmath m_{1/2}$. The  resulting relation between $M_1$ and $M_2$ is
$M_1=\frac{5}{3} tan^2\theta_W M_2\sim 0.5 M_2$,

\item
\mbox{\boldmath $m_{\tilde{\mathrm f}}$}, the sfermion masses.
Under the assumption of sfermion mass unification, \mbox{\boldmath${m_0}$} 
is the common sfermion mass at the GUT scale,

\item
the trilinear couplings \mbox{\boldmath$A_{\mathrm f}$} determining the 
mixing in the sfermion families.        
The third family trilinear couplings are the most relevant ones,   
\mbox{\boldmath $A_\tau, A_{\mathrm b}, A_{\mathrm t}$}.
\end{itemize}

Gaugino mass unification leads to  $m_{1/2} \simeq 1.2 M_2$ and 
to  the following approximate
relations between \MXC{1}, \MXN{1} and  the gluino mass ($m_{\tilde{g}}$):

\begin{itemize}
\item
in the  region where \XN{1} and \XPM{1} are {\it gauginos}  ($|\mu| >> M_1$),
\MXC{1} $\simeq $ \MXN{2} $\simeq$ 2 \MXN{1},
 $ m_{\tilde{g}} \simeq$   3.2 \MXC{1} and
\MXC{1}  $\simeq$   $M_2$,
\item
in the {\it higgsino} region ($|\mu| << M_1$), 
\MXC{1} $\simeq$ \MXN{2} $\simeq$  \MXN{1} $\simeq |\mu| $.\\ 
\end{itemize}

The relations between  chargino, neutralino and gluino masses and $|\mu|$
and $M_2$ are affected by radiative corrections of the order of 
2\%-20\% \cite{radcor}.
However, {\it only the relative relations between 
chargino, neutralino and gluino
masses  are important from the experimental point of view}, 
and here the corrections
are much smaller. For example, the relation
\MXC{1}/\MXN{1} $\simeq$ 2 in the gaugino
region, which is usually
exploited to set a limit on the LSP mass, 
receives the corrections only of the order of  2\%;
and the ratio $m_{\tilde{g}}/\MXC{1}$ $\simeq$  3.2 receives corrections of the
order of 6\%. Thus, for example, the limit \cite{susywg} on the chargino mass
of 103.5~\GeVcc\  set by LEP 
(valid for \msnu$>$300~\GeVcc, \mstauo$>$\MXC{1}, 
and for $M_2 \leqsim 200$~\GeVcc) 
can be safely translated to \MXN{1} $\geqsim$ 51~\GeVcc\ and  
$m_{\tilde{g}} \geqsim$~310~\GeVcc.
   
  If the  sleptons are heavy the chargino mass limit excludes regions 
in ($M_2,|\mu|$ ) plane  (see e.g. \cite{delphisusy}).
For  \tanb $\geqsim$, 2  $|\mu| \geqsim  $100 \GeVcc\ 
is excluded up to very
high values of $M_2$ (of the order of 1000~\GeVcc\ or more) while  
$M_2 \leqsim$ 100~\GeVcc\ is excluded for   
$|\mu| \geqsim  $100 \GeVcc. 

   Electroweak symmetry imposes the following relation between the masses of
the superpartners of the left-handed electron (\sell) and 
of the neutrino (\snu),

1) $\msell^2 = \msnu^2 + m_W^2 |cos2 \beta| $.

The assumption of sfermion mass unification relates masses
of the ``left-handed'' ($m_L$) and the ``right-handed'' 
($m_R$) ``light'' sfermions,  
``light'' squark  masses, and the gaugino mass
parameter  $M_2$. For example :

2)   $\msnu^2 $ $=$ $m_0^2+ 0.77M_2^2 -0.5\MZ ^2 |\cos 2\beta|$

3)   $m_L^2$ $=$ $m_0^2+ 0.77M_2^2  +(0.5 - sin^2\theta_W) \MZ ^2 |\cos 2\beta|$

4)   $m_R^2$ $=$ $m_0^2+ 0.22M_2^2 + sin^2\theta_W \MZ ^2 |\cos 2\beta|$
 
5)   $m_{d_L}$ $=$ $m_0^2 + 9 M_2^2 +(0.5 -1/3 sin^2\theta_W)\MZ ^2 |\cos 2\beta|$

Thus, for example,
$m_{d_L} \geqsim$~310~\GeVcc, if \MXC{1} $\geqsim$ 103.5~\GeVcc.\\
   Mixing between left and right states (present for superpartners of heavy
fermions) gives rise   to   off-diagonal  ``left-right'' mixing
terms in their mass matrices, which lead to a mass splitting between 
the lighter and the heavier state. 
 At the EW scale these terms are  proportional to
$m_{\tau}(A_\tau - \mu \tanb)$, $m_{b}(A_b - \mu \tanb)$ and
$m_{\mathrm t}(A_t - \mu / \tanb)$ for 
\stau, \sbq\ and \stq, respectively, where $A_{\tau}, A_b, A_t$ are free
parameters.
Therefore, for large $\mu$ this can give light stau and sbottom states
if \tanb\ is large, or a light stop for small \tanb.

     For large $m_A$, the lightest Higgs boson mass depends primarily on  
\tanb,  $m_{top}$ and the mixing in the stop sector $X_t$
(expressed here as $X_t=A_t - \mu / \tanb $), and this  dependence
is maintained whether any additional constraints on the  MSSM
are imposed or not. The top quark mass is presently known with the
uncerntainty (1$\sigma$) of around 5~\GeVcc\ \cite{pdg}, and
the resulting uncerntainty of the lightest Higgs boson mass calculation
is around 6.5~\GeVcc, 
as $\Delta m_{\hn}/m_{\hn} \simeq 2 \Delta m_{top}/m_{top}$.
It was
shown in \cite{carwig} that for a given \tanb\ and top mass, the maximal
$m_{\hn}$ occurs for   $X_t/m_{SUSY}= \sqrt{6}$.  Another, slightly lower
maximum occurs for   $X_t/m_{SUSY}= -\sqrt{6}$. $m_{SUSY}$ is typically
taken to be of the order of the  gluino mass, or of the 
diagonal terms in the squark mass matrices, and $m_{\hn}$ grows 
with $m_{SUSY}$.

   It should be noted that the off-diagonal terms in mass matrices of the third
family sparticles cannot be too big compared to the diagonal terms, in order
for a real solution for sparticle masses to exist. 
As diagonal terms grow  with  $m_0$ and $M_2$, for every
given value of the off-diagonal term 
a lower  limit is set on the corresponding 
combination of $m_0$ and $M_2$
\footnote{\protect{To avoid ``tachyonic'' mass solutions we must have:

$$m_{ll}+m_{rr}  > \sqrt{ (m_{ll}-m_{rr})^2 +4*m_{lr}^2 }$$

\noindent
where $m_{lr}$ is the off-diagonal mixing term, and $m_{ll},m_{rr}$
are the diagonal mass terms. For example, for the stop we have 
$m_{lr}=m_{top}X_t$ and,

$m_{ll} \simeq m_0^2 + 9M_2^2 +
m_{top}^2 + m_Z^2 cos2\beta (0.5-2/3sin^2\theta_W)$

$ m_{rr} \simeq m_0^2 + 8.3M_2^2 + m_{top}^2 + 
2/3 m_Z^2 cos2\beta sin^2\theta_W $

For an example value of $X_t=\sqrt{6}$ \TeVcc, 
the condition above sets a lower limit
on a combination of $m_0^2$  and   $M_2^2$:
${m_0}^2 + 8.5{M_2}^2 > 0.39$ \TeVcc\
Thus, if $m_0< 300$~\GeVcc\ we must have $M_2> 190$~\GeVcc. 
}}.

\noindent
\subsection{  {mSUGRA} }

In the minimal SUGRA model not only the sfermion masses, 
but also the Higgs masses 
$m_{H_1}$ and $m_{H_2}$, are assumed to unify to the common 
$m_0$ at the GUT scale. 
Then $m^2_{H_2}$ becomes negative at the  EW scale 
in most of the  parameter space, thus ensuring 
EW symmetry breaking.
 
   The additional requirements of the unification of the trilinear 
couplings to a common $A_0$
and the correct reproduction of the EW symmetry scale, which fixes the 
absolute value of $\mu$, defines the minimal gravity-broken MSSM
(mSUGRA).  The value   of $\mu^2$ can be determined minimising 
the Higgs potential and 
requiring  the right value of  $m_Z$. At tree level \cite{sugpeda}:
 
6) $\mu^2 =- 1/2m_Z^2+ \frac{ m^2_{H_1} -m^2_{H_2}tan^2\beta}{tan^2\beta -1}$

7)  $m^2_{H_1} \simeq  m^2_0 +0.5{m^2}_{1/2}$, 
$m^2_{H_2} \simeq -(0.275m^2_0+3.3{m^2}_{1/2})$

The parameter set is then reduced to  
$m_{1/2}, m_0, \tanb, A_0$ and the sign of $\mu$. \\
   In addition to the  mass relations listed in the previous subsection,  
$m_A$ can be related to $m_{1/2}$ ($M_2$), $m_0$ and Yukawa 
coupling of the top quark.
The stop
mixing parameter can be expressed (approximately) 
as $A_t =0.25 A_0 -2 m_{1/2}$.  For low \tanb,   
$m_A^2$ $\simeq$ $m_0^2 + 3m_{1/2}^2 -m_Z^2$. As $m_{\hn}$ grows
with $m_A$ and $A_t$ (see section \protect{\ref{sec:intro}}), 
Higgs searches  can be used to set a limit on  $m_{1/2}$ ($M_2$)
which depends on \tanb, $A_0$, and $m_{top}$.
The
lightest Higgs mass can thus be related to 
$m_{1/2}$ ($M_2$), and the experimental
limit on it can be used to set  
limits 
on the masses of 
(for example) the lightest
chargino and the lightest neutralino
dependent on
\tanb, $A_0$ and $m_{top}$.

\section{LEP results}
\label{sec:search}

In years  1995-2000, the Aleph, DELPHI, L3 and OPAL experiments at LEP 
collected an integrated
luminosity of more than  2000~\pbi\ at  centre-of-mass energies ranging 
from 130~GeV to 208~GeV.
These data have been analysed to search for the sfermions, 
charginos, neutralinos and Higgs bosons predicted by supersymmetric 
models 
\cite{higdelphi,delphisusy,opalsusy,l3susy,alephsusy,higaleph,higopal,higl3}.

\subsection{Searches for charginos and neutralinos  }
\label{sec:searchar}

\begin{figure}[!hbt]
\begin{center}
\begin{tabular}{cc}
\resizebox{80mm}{!}{\includegraphics{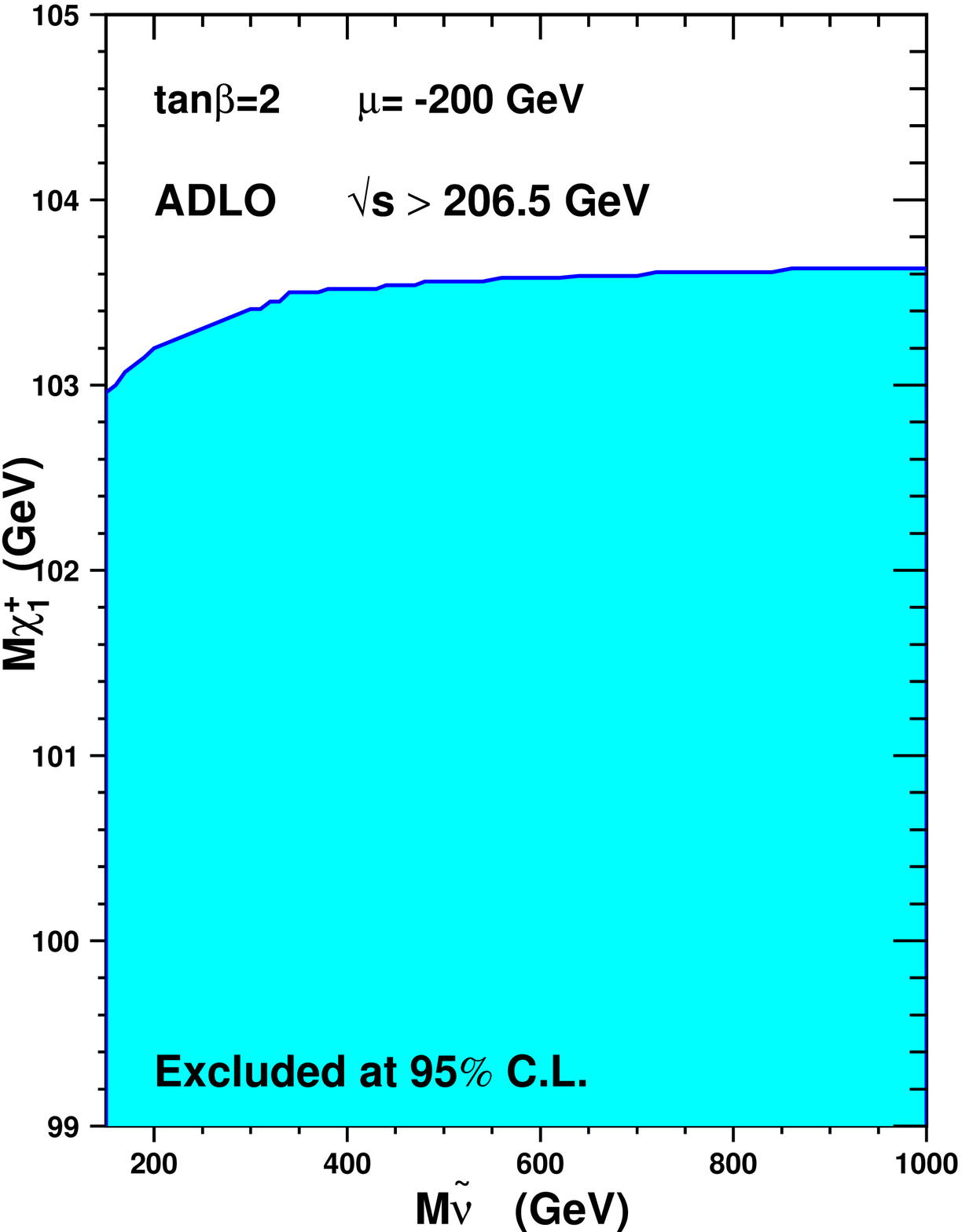}}&
\resizebox{80mm}{!}{\includegraphics{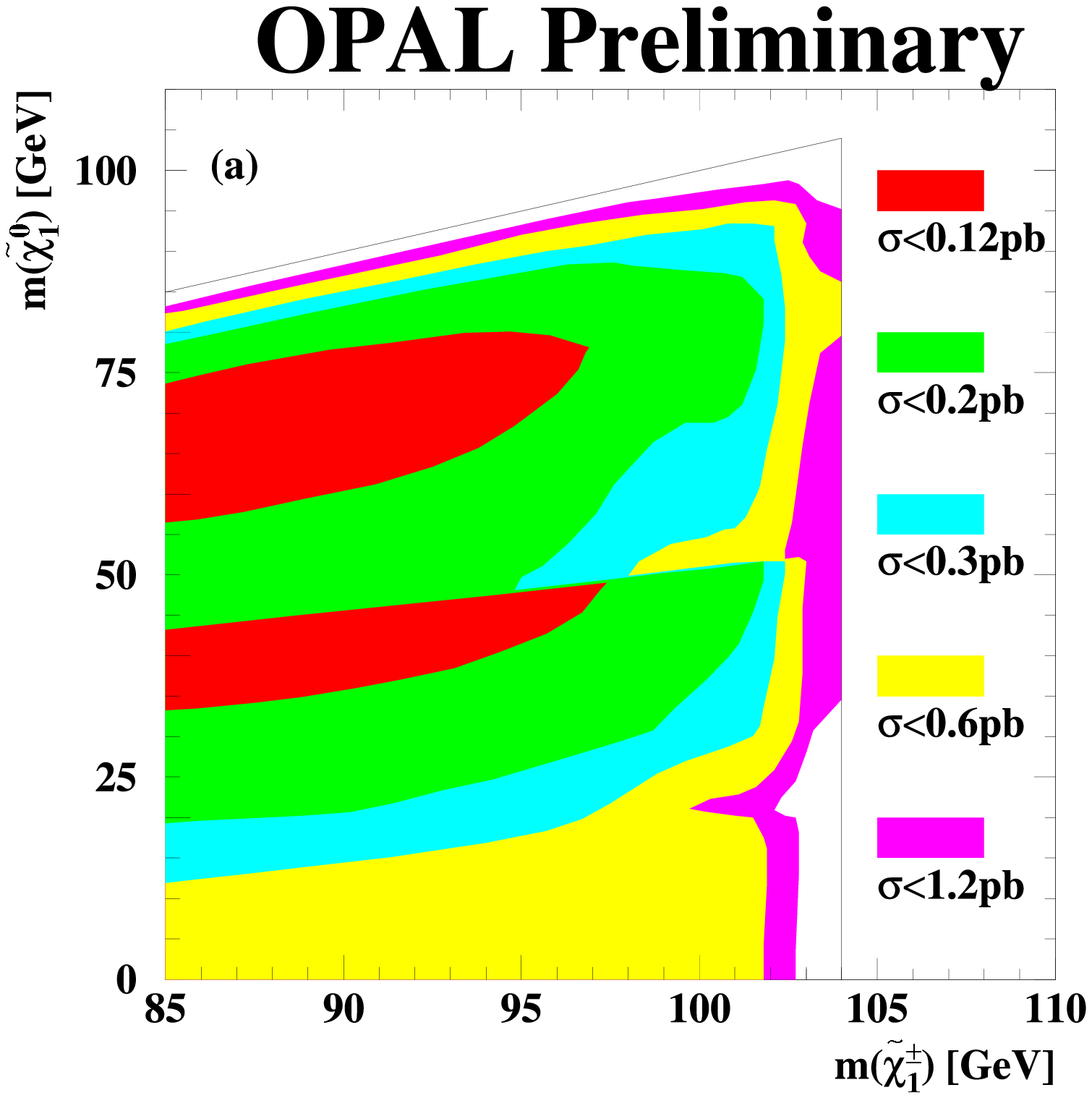}}\\
\end{tabular}
\caption[MSSM limits in ($\mu$,$M_2$) plane]{
Left hand side: Limit on the chargino mass at 95~\% confidence level,
 resulting obtained by the LEP SUSY working group (see text). 
The limit is  valid for the decay channel $\XPM{1} \to \XN{1} W^*$
Right hand side:
Limits on the chargino production
cross-section in the   ($\MXN{1}, \MXC{1}$)  
plane, at 95~\% confidence level, resulting from Opal searches. 
The limits are valid for the decay channel $\XPM{1} \to \XN{1} W^*$
}
\label{fig:charlim}
\end{center}
\end{figure}

After the  Higgs \cite{rosy}, charginos were the most important
SUSY discovery channel at LEP. Unless there is a light sneutrino 
(in the gaugino region
the chargino production cross-section can be quite small due to
the negative interference between the t-channel sneutrino exchange diagram
and the s-channel $Z/\gamma$ exchange diagram. 
Higgsino-type charginos do not
couple to the sneutrino.), the  chargino pair production
cross-section is predicted to be large  if $\MXC{1}< \sqrt{s}/2$. 
A lower  limit
on the chargino mass of 103.5~\GeVcc\ was set
\cite{susywg}, shown on figure \ref{fig:charlim}
assuming 100\% branching fraction
to  the  decay mode
$\XPM{1} \to \XN{1} W^*$. Although this limit is ``earmarked'' to be
set only in one MSSM point, it is valid as long as the chargino decays as
above. \\

    Cross-section limits for chargino pair-production 
were set (see figure \ref{fig:charlim}).
They depend primarily on the difference between the mass of the chargino and an
undetectable sparticle it decays to (e.g. \XN{1} or \snu ).   
Chargino pair production with cross-section larger than  0.1-0.2~pb 
(corresponding  to 
$\sqrt{s} \sim$ 205 GeV, the average energy of the year 2000 data)
is  excluded 
 for $\DM~>~20~\GeVcc$ \cite{opalsusy,delamsb}, where
\DM~=~\MXC{1}~$-$~\MXN{1} or \DM~=~\MXC{1}~$-$~\msnu.
If these limits are combined, a  chargino production cross-section
above   0.05~pb-0.1~pb can be excluded.  The limit on the chargino mass
of \MXC{1} $\geqsim$ 100~\GeVcc\ can be set  for the light
sneutrino as well,  as long as  $\DM~\geqsim~10~\GeVcc$.
Alas, no official LEP combination exists for the chargino decaying
to the sneutrino and a lepton.

   If sfermion mass unification is assumed, searches for \selr\ can be used  
to set a lower limit on the sneutrino mass, and thus on the chargino mass
in the case of a light sneutrino and $\DM~<~10~\GeVcc$. Moreover,
if \sell\ and \selr\ are light, neutralino production in the
gaugino region is enhanced  ( experimentally observable
neutralino production (for example \XN{1}\XN{2}) has quite
large cross-section in the higgsino region as higgsinos
couple directly to \Zn. However, in the gaugino region there is
no tree-level coupling of \XN{1} to \Zn, and $\ee \to \XN{1}\XN{2}$
can only be mediated via t-channel selectron exchange ), and
neutralino searches 
set an indirect limit on the sneutrino mass in some regions of the parameter
space.

   Another "blind-spot" in chargino 
searches arises when the \stauo\ is light and close in mass
to  the $\XN{1}$ \cite{delphisusy,susywg}. Chargino decays 
$\XPM{1} \to \stauo\ \nu$ with $\stauo \to \XN{1} \tau$ then
dominate, and lead to an ``invisible'' final state; but   
the search for neutralino production  can be used \cite{delphisusy,susywg} 
in this
case.
If neutralinos decay  via light stau states and 
$\mstau$ is close to $\MXN{1}$,
$\XN{1}\XN{2} $  production  with 
\XN{2}~$\to$~\stau$\tau$ and \stau~$\to$~\XN{1}$\tau$ leads to
only one $\tau$ visible in the detector;  nevertheless 
limits  on the cross-section times branching ratio 
are of the order of 0.1-0.4 pb \cite{delamsb}. 
The search for  $\XN{2}\XN{2}$ in the same region 
reaches  a sensitivity of 0.06 pb \cite{delphisusy}.  
In the CMSSM with nUHP, the region 
in ($M_2,\mu, m_0$) space where
the stau is degenerate in mass with the LSP depends on mixing parameters: 
$A_{\tau}$, 
and  $A_{b}$,$A_{t}$. It is possible to find  configurations of mixing
parameters  (typically
with $|\mu|$ few times larger than $M_2$ and $m_0$)  
such that the  stau is light and close in mass to \XN{1} while
the selectrons are heavy, rendering the neutralino cross-section small.
However, the chargino production cross-section is large in this case,
and this region can be 
explored  by the search for $\XPM{1}\XPM{1} \gamma$  production 
\cite{delphisusy,chadege,alephnote}  where the photon arises from 
initial state radiation  and is detected together with 
a few low energy
tracks originating from \XN{2}~$\to$~\stau$\tau$ and \stau~$\to$~\XN{1}$\tau$ 
decay chain.

   In mSUGRA, $|\mu|^2$ is in the range
3.3~$m^2_{1/2}$-0.5$m_Z^2$~$ < \mu^2 < $~$m^2_{0}+3.8 m^2_{1/2}$ 
for \tanb$>$2 and
and light stau cannot be degenerate with neutralino
for large $m_0$. Thus  neutralino searches   
set  a limit on  the chargino mass  for small $\mstauo -\MXN{1}$ which 
is close to the one obtained for  heavy sleptons (around 103~\GeVcc).

   It is perhaps worth mentioning that, because  in the
higgsino region ($M_1>>|\mu|$)  the  $\XN{1} \XN{2}$ production
cross-sections at LEP are large,
$\XN{1} \XN{2}$ production can be excluded nearly up to the
kinematic limit as long as $\MXN{1}$ is not too close to $\MXN{2}$
($M_2 \leqsim 1500 $~\GeVcc\ in the constrained MSSM). 
For
$200 < M_2 < 1500 $~\GeVcc\ 
a lower limit on the LSP mass of 70~\GeVcc\ 
was set by DELPHI \cite{lim189},
using the data collected
at $\sqrt{s}$= 189~GeV. In the  constrained MSSM   
the mass difference between the 
lightest chargino and the lightest neutralino
is less  than  3~\GeVcc\ for $M_2 \geqsim 1500$~\GeVcc. 
A lower limit 
on the $\MXC{1}$ of around 92~\GeVcc\ was set in this region
by LEP SUSY working group,\cite{susywg}, 
implying a similar lower limit on the mass of the lightest
neutralino.   

\noindent
\subsection{Searches for Sleptons and Squarks}

Pair-produced selectrons and muons with the 
typical decay modes,
$\sle \to \XN{1} \ell$, have been searched for by all LEP
collaborations.
These searches exclude slepton pair production
with a cross-section above (0.02-0.1)~pb
depending on the
neutralino mass and on the slepton mass, 
assuming
100\% branching fraction to the above decay mode.
With this assumptions, right-handed smuons (selectrons) 
lighter than around 
96 (99)~\GeVcc\ can be excluded,
provided $\msmur(\mselr) -\MXN{1} \geqsim 20$~\GeVcc\ and that the
selectron pair production cross-section is as for \tanb=2, $\mu$=$-$200.\\
      For the minimal coupling to $Z/\gamma$ and sufficiently
large  \DM~=~\mstauo~$-$~\MXN{1} $>$ 15 \GeVcc, 
\mstauo $\leqsim$ 85~\GeVcc\ can be excluded, while the lower limit 
on the mass of the stable stau is close to 97~\GeVcc.\\
      It should be noted that while 
selectron production cross-section depends on the neutralino mass
and composition, the smuon and stau production cross-section
depends only on the sparticle handness and mass, thus the limit
presented here is valid as long as smuons(staus) decay as above.

\begin{figure}[!hbt]
\begin{center}
\vskip -0.5 cm
\begin{tabular}{cc}
\resizebox{80mm}{!}{\includegraphics{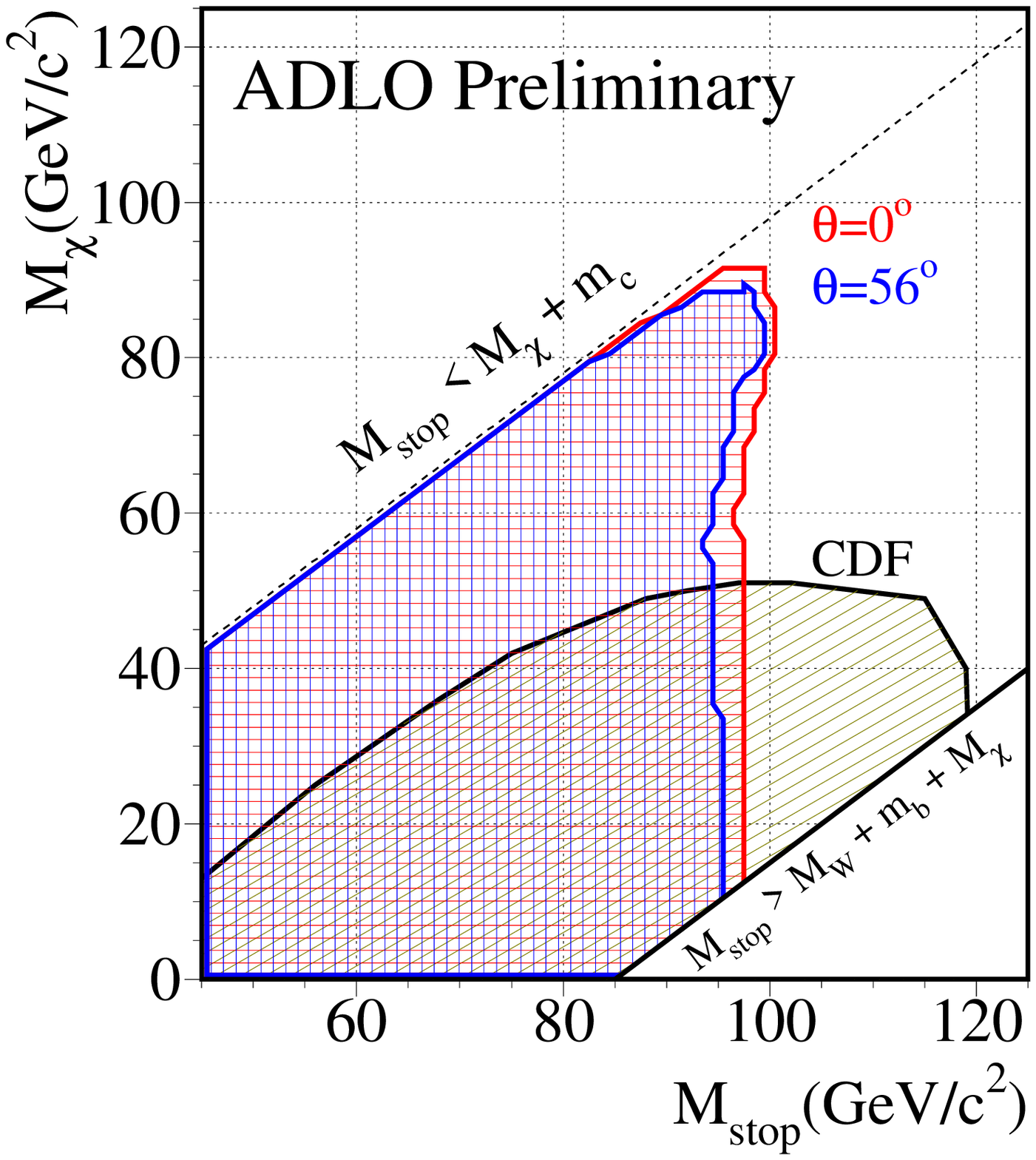}}&
\resizebox{80mm}{!}{\includegraphics{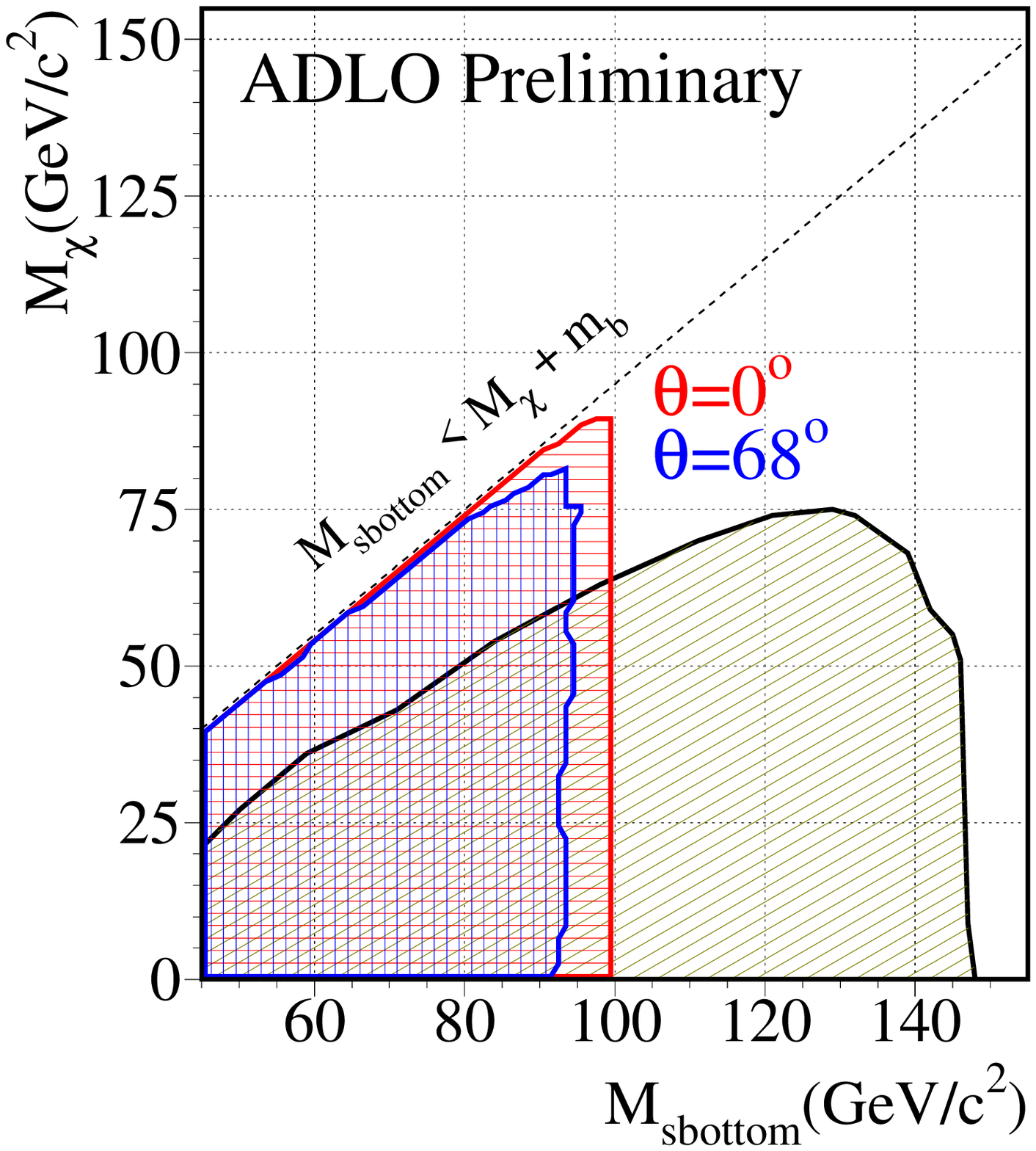}}\\
\end{tabular}
\caption[MSSM limits in ($\mu$,$M_2$) plane]{
Exluded ranges in  ($m_{\stq_{1}}$,\MXN{1})  and ($m_{\sbq_{1}}$,\MXN{1})
planes, at 95~\% confidence level, resulting from the combined
Aleph, Delphi, Opal and L3 searches. 
The hatched 
shading is excluded by the CDF collaboration, assuming mass degeneracy
between the lighter and the heavier stop (sbottom) states. 
(see text). 
}
\label{fig:stopsbot}
\end{center}
\end{figure}

     The results of the searches for  sbottom ($\sbq $) and
stop ($\stq $) 
were combined by the LEP SUSY working group.
The typical decay modes 
$\tilde{t} \to \XN{1}  c$ and $\tilde{b} \to \XN{1}  b$
have been searched for.
These searches exclude squark pair production
with a cross-section   above (0.05-0.1)~pb
depending on the
neutralino  and on the squark masses,
assuming
100\% branching fraction to the above decay modes. 
For the minimal coupling to
$Z/\gamma$  and for  
\DM~=~\mstq(\msbq)~$-$~\MXN{1} $>$ 15 \GeVcc, 
the \stq (\sbq) with  mass below
 95~(93)~\GeVcc\ is then excluded, as it can 
be seen on figure \ref{fig:stopsbot}, \cite{susywg}.

\section{Limits in CMSSM and mSUGRA scenario}
\label{sec:limits}                                           

The searches described in the previous section were used to set limits
on sparticles masses in the CMSSM with non universal Higgs parameters and 
in mSUGRA. 
{\it Limits presented in this section are valid in the R-parity conserving
scenario and in all R-parity violating scenarios where a  chargino limit
of 103~{\boldmath \GeVcc }
or more can be set by LEP experiments.}

\subsection{Limits in the CMSSM with nUHP}

Higgs boson searches and chargino searches set limits in this scenario. 
"Holes" which arise  in chargino searches in the 
R-parity conserving scenario
are covered
by selectron, neutralino, Higgs and squark searches.  Limits presented
in this section are for $m_A$ $\le$ 2000~\GeVcc. \\


\noindent
 {\it{Limits on the  mass of the lightest neutralino}}

Efforts of LEP collaborations went into covering various blind spots
in the chargino searches, in order to set an ``absolute'' neutralino mass
limit (within the CMSSM). As explained below, the limit is set in one
of the two quite highly fine-tuned blind spots: chargino-sneutrino mass degeneracy
with \DM $\leqsim$ 3 \GeVcc\ and stau-neutralino mass degeneracy with
similar \DM. As none of these situations is likely to occur it is probably
worth asking, what would be the neutralino mass limit if both
of these degeneracies are avoided. The answer will be given at the end of this
subsection.

\begin{figure}[!hbt]
\begin{center}
\vskip 0.1 cm
\begin{tabular}{cc}
\resizebox{80mm}{!}{\includegraphics{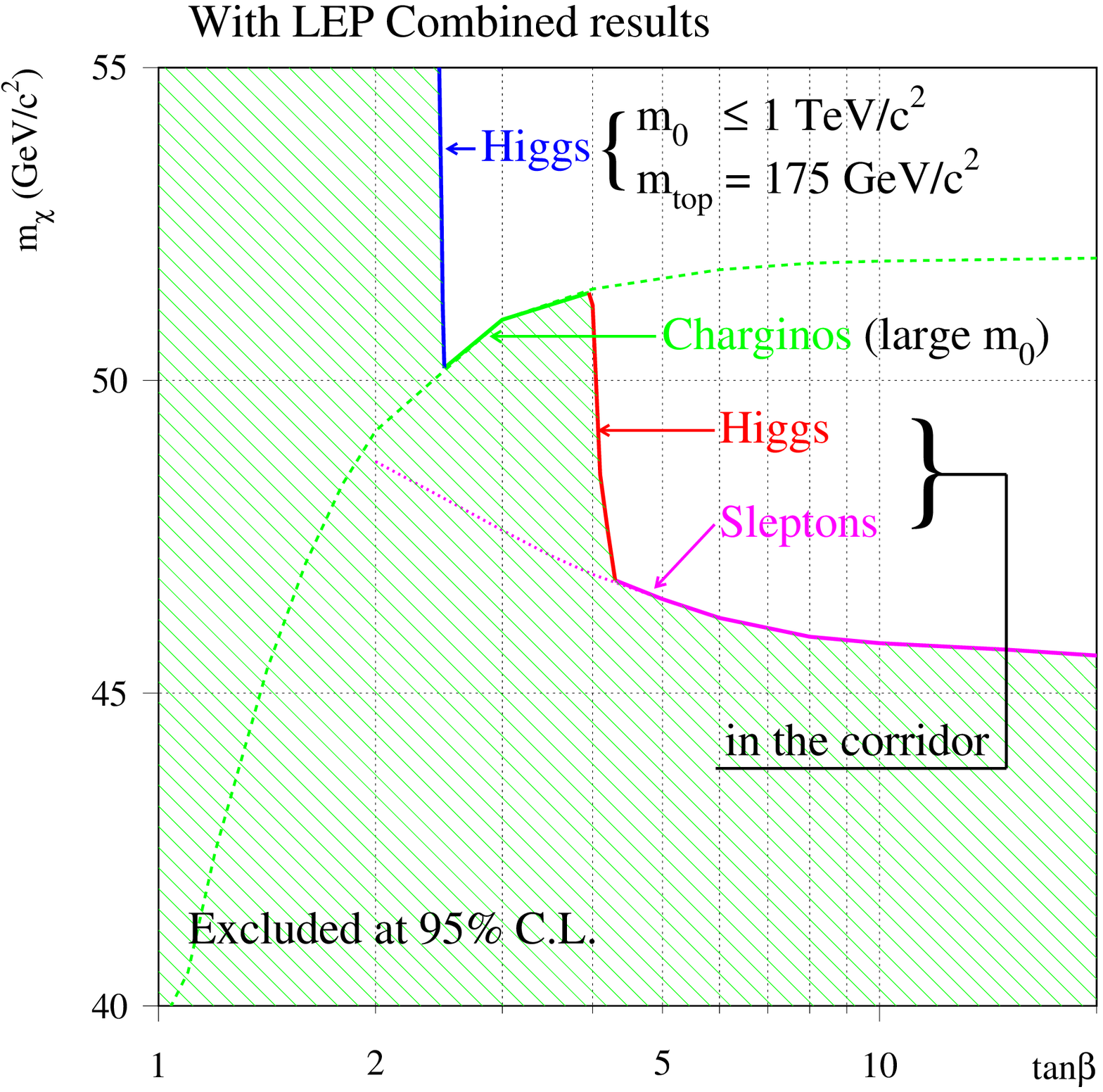}}
\resizebox{80mm}{!}{\includegraphics{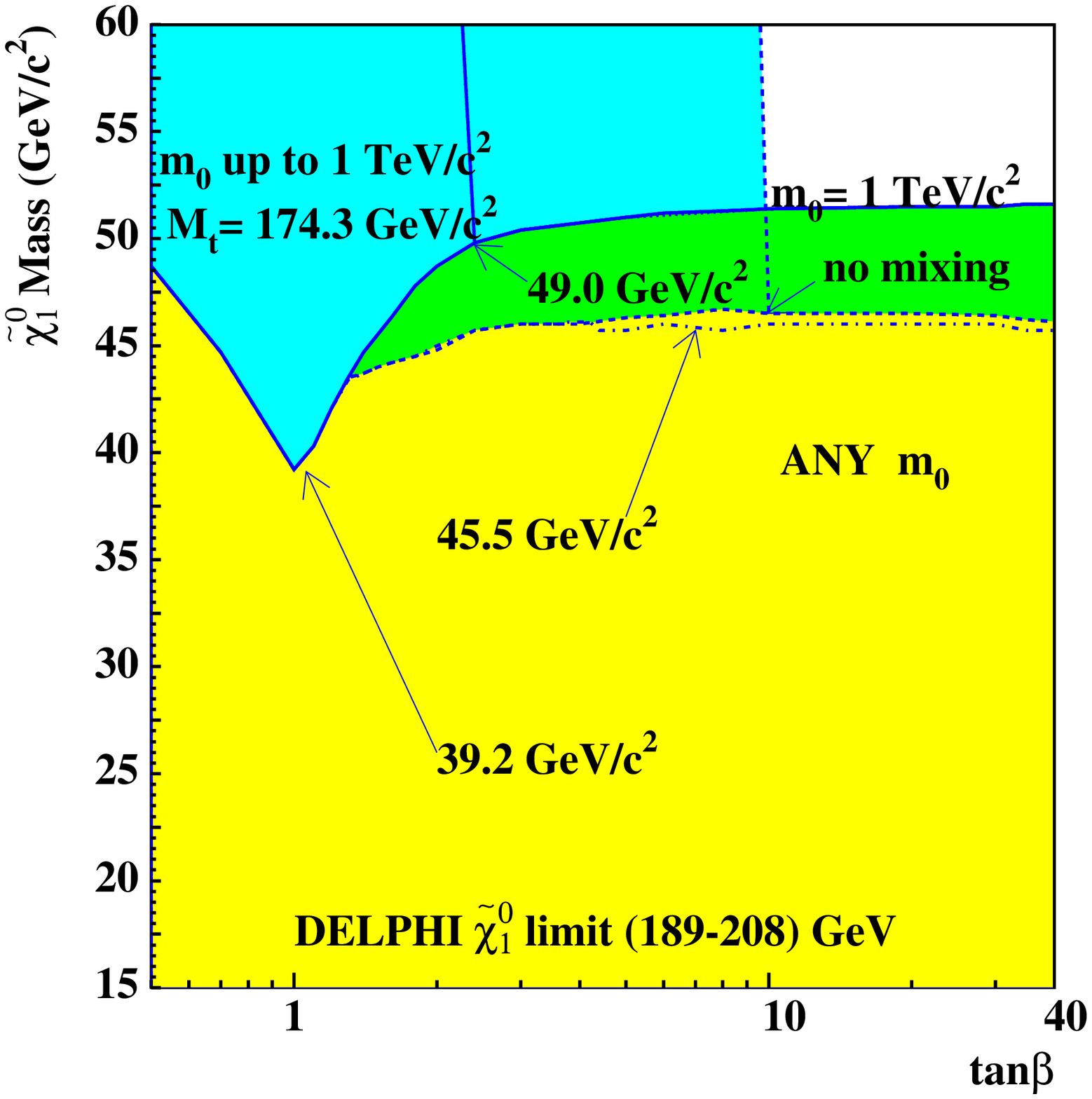}}
\end{tabular}
\caption[MSSM limits in ($\mu$,$M_2$) plane]{
The lower limit at 95~\% confidence level on the mass of the lightest
neutralino, \XN{1} as a function of
\tanb\ assuming a stable \XN{1}.
Left hand side: 
The limit is set by the ``LEP combined'' Higgs, chargino and selectron
searches assuming gaugino and sfermion mass unification. It is
valid for small (no) mixing in the stau sector.
Right hand side, DELPHI:
The solid curve shows the limit obtained for 
$m_0$~=1000~\GeVcc, the dashed curve shows the limit obtained
allowing for  any $m_0$ assuming that there is no mixing in the
third family
($A_{\tau}=\mu\tanb$, $A_{b}=\mu\tanb$, $A_{t}=\mu/\tanb$),
and the dash-dotted curve shows the limit obtained for
any $m_0$ allowing for the mixing with 
$A_{\tau}$=$A_{b}$=$A_{t}$=0. 
The steep solid (dashed) curve  shows the effect of the
searches for the Higgs boson
for the maximal $M_{\hn}$ scenario (no mixing scenario),  
$m_0$ $\le$ 1000~\GeVcc\ and
$M_t$= 174.3~\GeVcc, which amounts to excluding the region of 
$\tanb<2.36 (9.7)$ }

\label{fig:lspdelphi}
\end{center}
\end{figure}

  The effect of various searches is illustrated 
on figure \ref{fig:lspdelphi} showing the LSP mass
limit set by the  Higgs and SUSY, as a function of \tanb. 

The mixing in the stop sector was of the form, $A_{t}-\mu/\tanb$), while
it was assumed that mixing in the sbottom and stau sector
is negligible.  Mixing in the stop sector was tuned to maximize the
$m_h$ for any given $M_2$, while avoiding the
tachyonic stop. Limit on the $m_h$ set by LEP
at low \tanb\ $<$ 6 sets a limit on $M_2$ for \tanb $<$ 2.4,
and for \tanb $<$ 4  a limit on the combination of $m_0$ and
$M_2$ is set  which excludes the region of chargino-sneutrino
degeneracy (where chargino searches are ineffective).
At higher \tanb\ this region is covered by the slepton
searches (primarily \selr), and the value of the
neutralino mass limit at large \tanb\ depends directly
on the value of the selectron mass limit for \MXN{1} simeq 45~\GeVcc.
The details of the limit derivation
can be found in \cite{susywg}. ``LEP combined'' Higgs, chargino and
selectron searches were used.\\
     DELPHI has obtained a similar limit assuming that mixing 
in the third family is of the form
($A_{\tau}-\mu\tanb$, $A_{b}-\mu\tanb$, $A_{t}-\mu/\tanb$, with
$A_{b}=A_{\tau}=0$ and $A_t$ in the range (0, $\pm$ maximal mixing),
see figure \ref{fig:lspdelphi}.\\
   If \mstauo=\MXN{1} is allowed by the large mixing
in the stau sector (the dotted line) the limit drops
at high \tanb\ to 45.5~\GeVcc, because  another  hole  in chargino and 
stau searches develops.
This "hole" is partially covered by neutralino and ``degenerate'' 
chargino searches \cite{delphisusy,alephnote}.\\   
     As before, the limit for ``any $m_0$'' with no mixing a drops
at \tanb$>$10 due
to the "hole" in chargino searches, where the chargino is close in mass
to the sneutrino. The "hole" is partially covered by selectron 
and neutralino
searches, and by the Higgs searches, which, in ``no-mixing''
scenario exclude \tanb\ $<$ 9.7.  The \tanb\ region excluded by Higgs searches
both in no-mixing, and in maximal mixing scenario depends on the mass
of the top quark and on the details of the Higgs mass calculations.\\   
   However, both in ``mixing'' and in ``no-mixing `` scenario the neutralino
mass limit is set at large \tanb, where the Higgs search has no effect.
While in the no-mixing scenario it is determined by the selectron mass
limit, in the mixing scenario it depends on the stau mixing model, and
on the interplay between chargino and neutralino searches with
\mstauo=\MXN{1}. \\    
   ALEPH and DELPHI have presented limits on the neutralino mass, in which stau
mixing is independent on stop or sbottom mixing, and has an arbitrary
$\mu$ dependence. In such a case stau can be made degenerate with the lightest
neutralino for any value of $M_2$ and $m_0$. As the sensitivity of neutralino
(degenerate chargino) searches used in  this case drops (grows) with $m_0$
the limit is set on the ridge of the exclusion from chargino and neutralino
searches. 
The limit is close to 41~\GeVcc\ and   represents the most
conservative scenario.

What would be the neutralino mass limit, if the ``sneutrino'' hole
and ``stau'' hole were avoided? The most pesymistic case is 
still the ``light sneutrino scenario'', which renders chargino production
cross-section small and enhances invisible decays of \XN{2}. For the
sneutrino lighter than the chargino and lighter than 65 \GeVcc\ DELPHI
alone sets a limit on the chargino mass of around 100 \GeVcc, independent
of \tanb\  
( see \cite{delamsb}). Similar limit can be set for the sneutrino 
mass just above the chargino mass. If  the data from all LEP experiments
are combined the gap in the sneutrino masses 65-100~\GeVcc\ is 
closed down to \DM~=~\MXC{1}~$-$~\msnu $\sim$  10~\GeVcc,
resulting in the chargino mass limit
$\sim$ 100~\GeVcc, and neutralino mass limit of
$\sim$ 50~\GeVcc (valid as long as  \DM~=~\MXC{1}~$-$~\msnu $>$ 10 \GeVcc).
Alas, no official combination of this decay channel was performed so
far. It is also interesting to note, that outside the stau and sneutrino
hole the limit on the neutralino mass can be set  which is 
{\it independent of the sfermion unification assumption}.\\

\noindent
{\it{Limits on the  masses of other sparticles}}

Limits on the masses of other sparticles can be set within the CMSSM,
which do not depend on a specific decay channel, but
take into account all decay channels appearing in the model. 
Also limits on the masses of sparticles, which are not directly visible
or produced at LEP can be set, due to their relations to the  masses of
observable sparticles (see section \ref{sec:cmssm}). 

\begin{figure}[!hbt]
\begin{center}
\vskip 0.5 cm
\begin{tabular}{cc}
\resizebox{80mm}{!}{\includegraphics{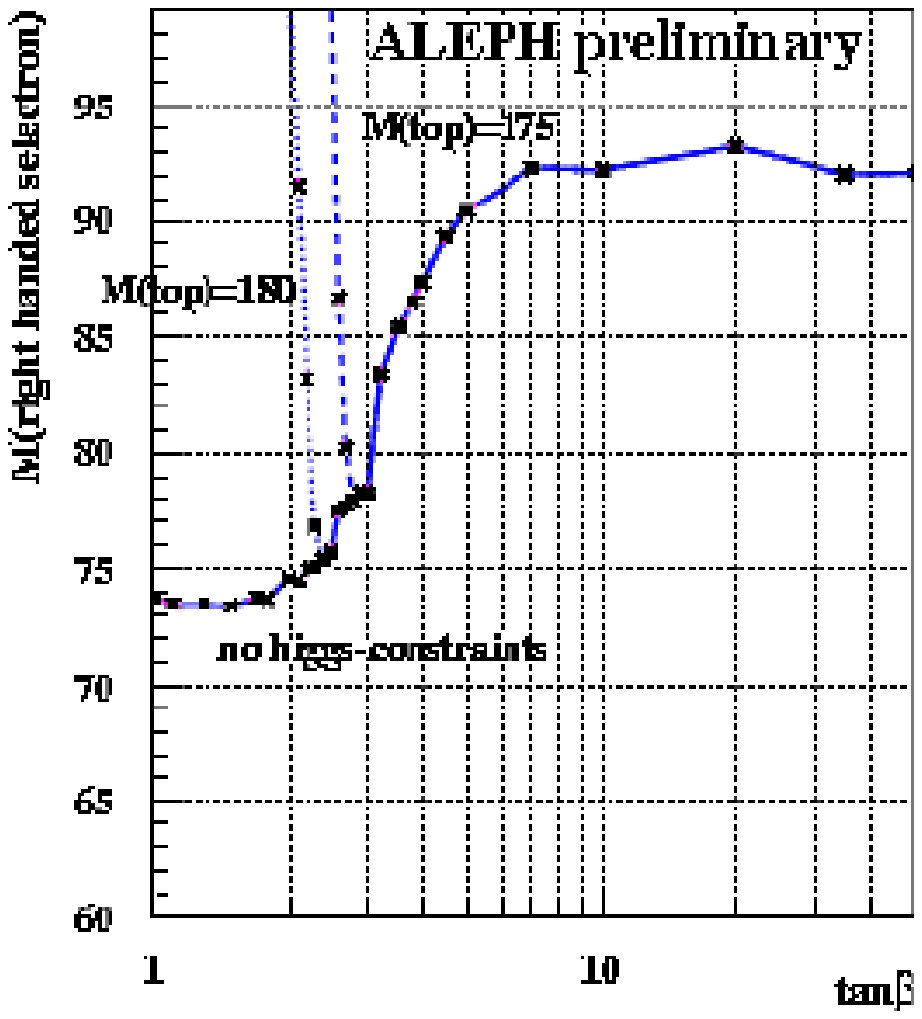}}&
\resizebox{80mm}{!}{\includegraphics{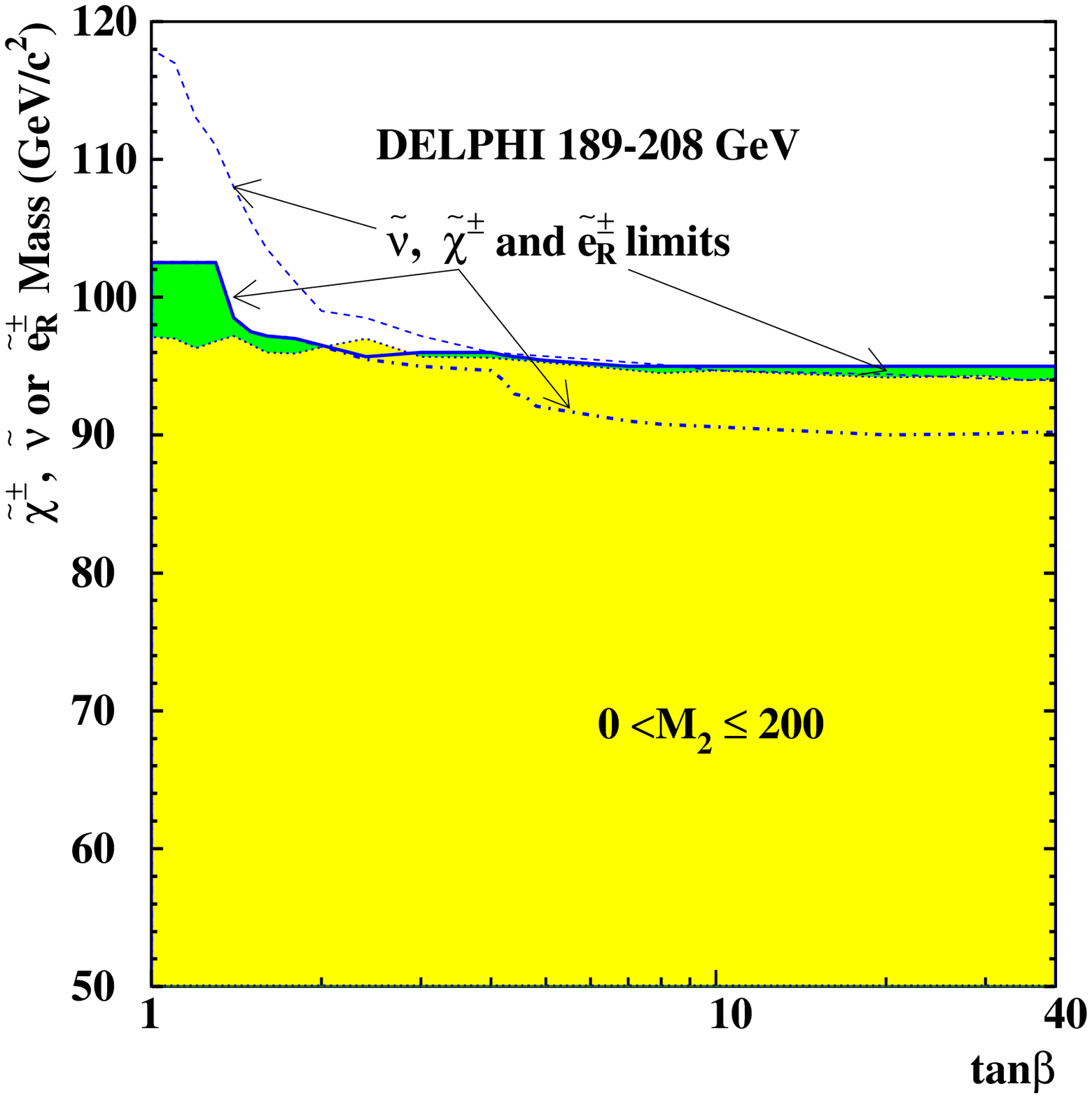}}\\
\end{tabular}

\caption[MSSM limits in ($\mu$,$M_2$) plane]{
Left hand side:The minimum
\selr\ mass in the CMSSM from Aleph.
The full line shows the limit without the constraint
from Higgs search. 
Right hand side, DELPHI: the minimum
sneutrino mass  (dark shading and
dashed curve) allowed by the slepton and neutralino
searches, as a function
of \tanb, together with the limits on the chargino mass (solid
curve and dash-dotted curve), and the \selr\ mass (dotted curve and
light shading). The chargino mass limit indicated by the solid curve and
the sneutrino and selectron mass limits  were
obtained assuming no mass splitting in the third sfermion family  
($A_{\tau}-\mu\tanb$=0 in particular). The chargino mass
limit is valid for $M_2 \leqsim $ 1500~\GeVcc. 
The selectron mass limit is valid for 
$\mselr- \MXN{1} > 10 $ \GeVcc.
The chargino mass limit indicated with the dash-dotted curve was obtained
allowing for mass splitting in the third sfermion family, 
with $A_{\tau}=A_{b}=A_{t}$=0.
}
\label{fig:seldelphi}
\end{center}
\end{figure}

   Aleph and DELPHI have set limits on the \msnu\ and \mselr\ which are valid
within the CMSSM (see figures  \ref{fig:seldelphi}).

Aleph limit is also valid for the mass configurations where the selectron
is degenerate with the lightest neutralino,
which occur at small \tanb. At higher \tanb\
the \mselr\ (\msnu) mass limit is close to 92-94 \GeVcc\ (88-94~\GeVcc). 
These limits were set for no mixing in the stau sector, which represents
in this case the most conservative scenario.

\begin{figure}[hbt]
\begin{center}

\resizebox{120mm}{!}{\includegraphics{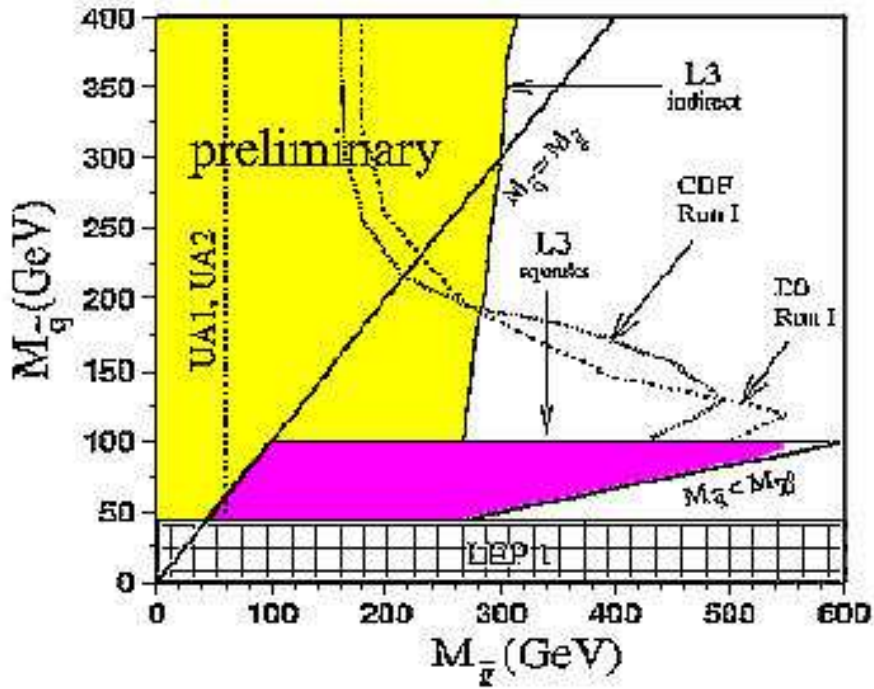}}

\caption[MSSM limits in ($\mu$,$M_2$) plane]{
The minimum
gluino  mass in the CMSSM from L3 (light shading). Dark shading
shows the limit on the squark mass, with the assumption that
all three squarks are mass-degenerate.
}
\label{fig:l3squark}
\end{center}
\end{figure}

Limits on the masses of the partners of light quarks and on the gluino mass
can be set as well, due to their relation to the chargino and
slepton masses (see section \ref{sec:cmssm} and \cite{mytevpaper}).\\ 
   L3 collaboration \cite{macpherson}  has used chargino-gluino 
mass relation to set an
indirect lower limit on the gluino mass of $\sim$ 300~\GeVcc\ 
(see figure \ref{fig:l3squark}). A lower mass limit on  ``light'' squarks
was set as well, exploiting the stop and sbottom searches, and assuming
that all  squarks are mass-degenerate.

   The relation between the chargino ($M_2$), 
slepton ($m_0$, $M_2$) and light squark masses ($m_0$, $M_2$)   was
exploited in \cite{mytevpaper} to set an indirect 
limit on the  $\tilde d$ mass of $\sim$~300~\GeVcc.\\

\subsection{Limits in the mSUGRA scenario}

Limits on the mSUGRA model for $A_0=0$  were discussed
in detail in ex. \cite{ellishigtb,roszk}. 
The Higgs search plays a major r\^ole
in setting these limits, and the value of $m_{\hn}$ depends crucially 
on $A_t \simeq 0.25 A_0 - 2m_{1/2}$, as it was
noted in for example \cite{mytevpaper, wimheavy}. A range of  $A_0$
was studied by the LEP SUSY working group \cite{susywg,macpherson}, and the
dependence of the results on the value of the top mass
was discussed. Even with the top mass fixed there is an
additional dependence of the exclusion on  the accuracy
of the $m_{\hn}$ calculations.

\begin{figure}[hbt]
\begin{center}
\begin{tabular}{cc}
\resizebox{75mm}{!}{\includegraphics{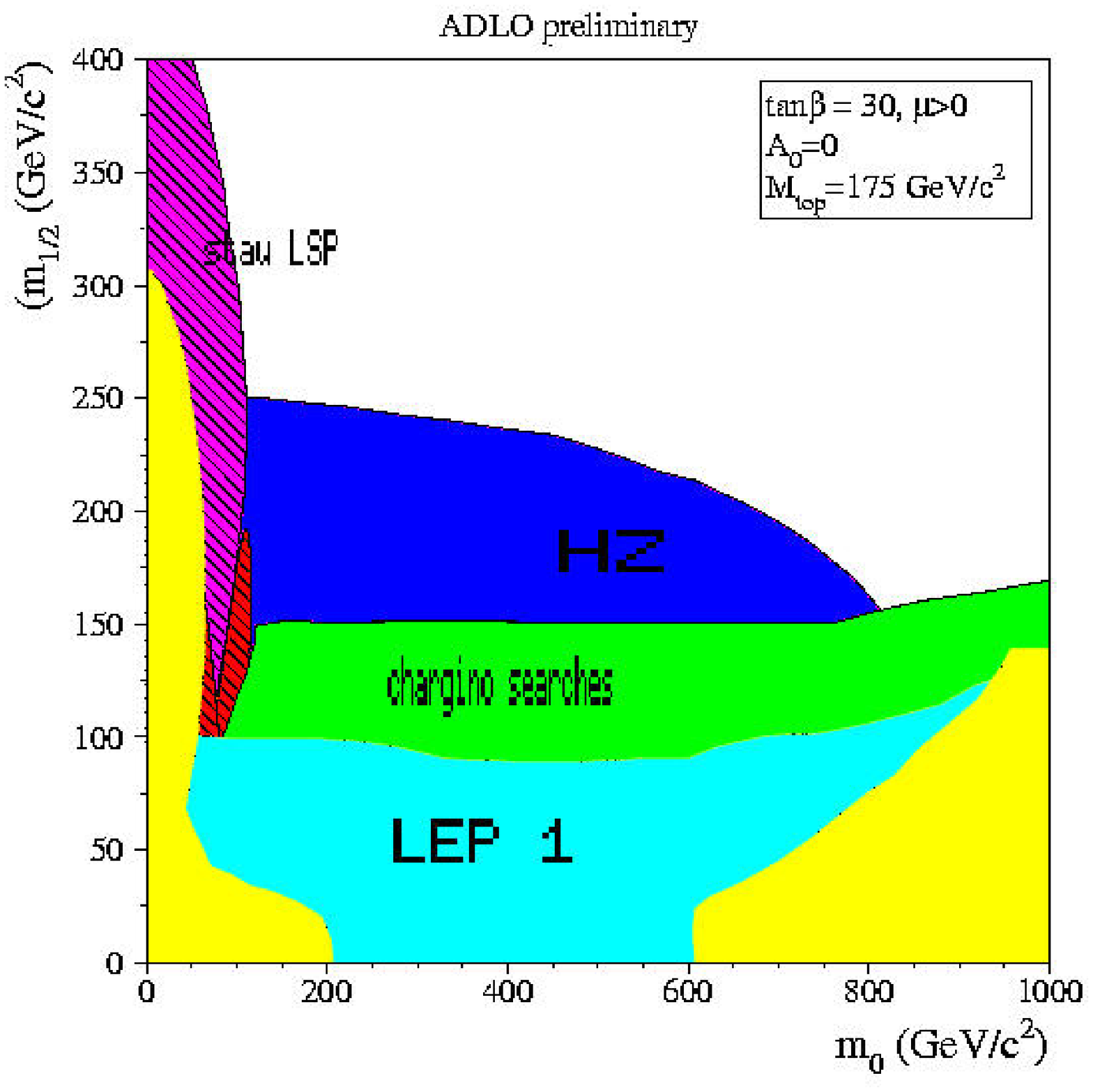}}&
\resizebox{80mm}{!}{\includegraphics{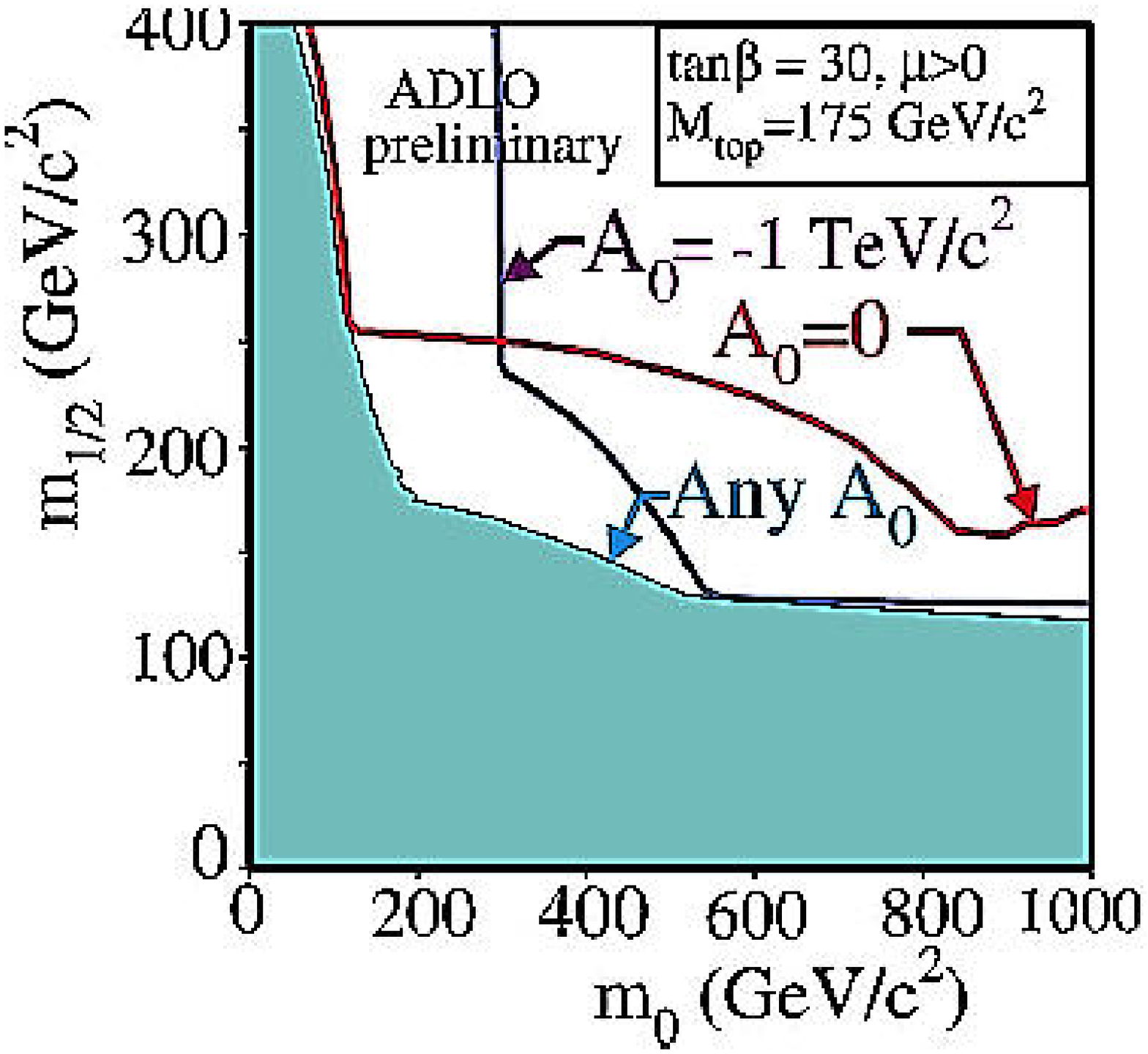}}\\
\resizebox{75mm}{!}{\includegraphics{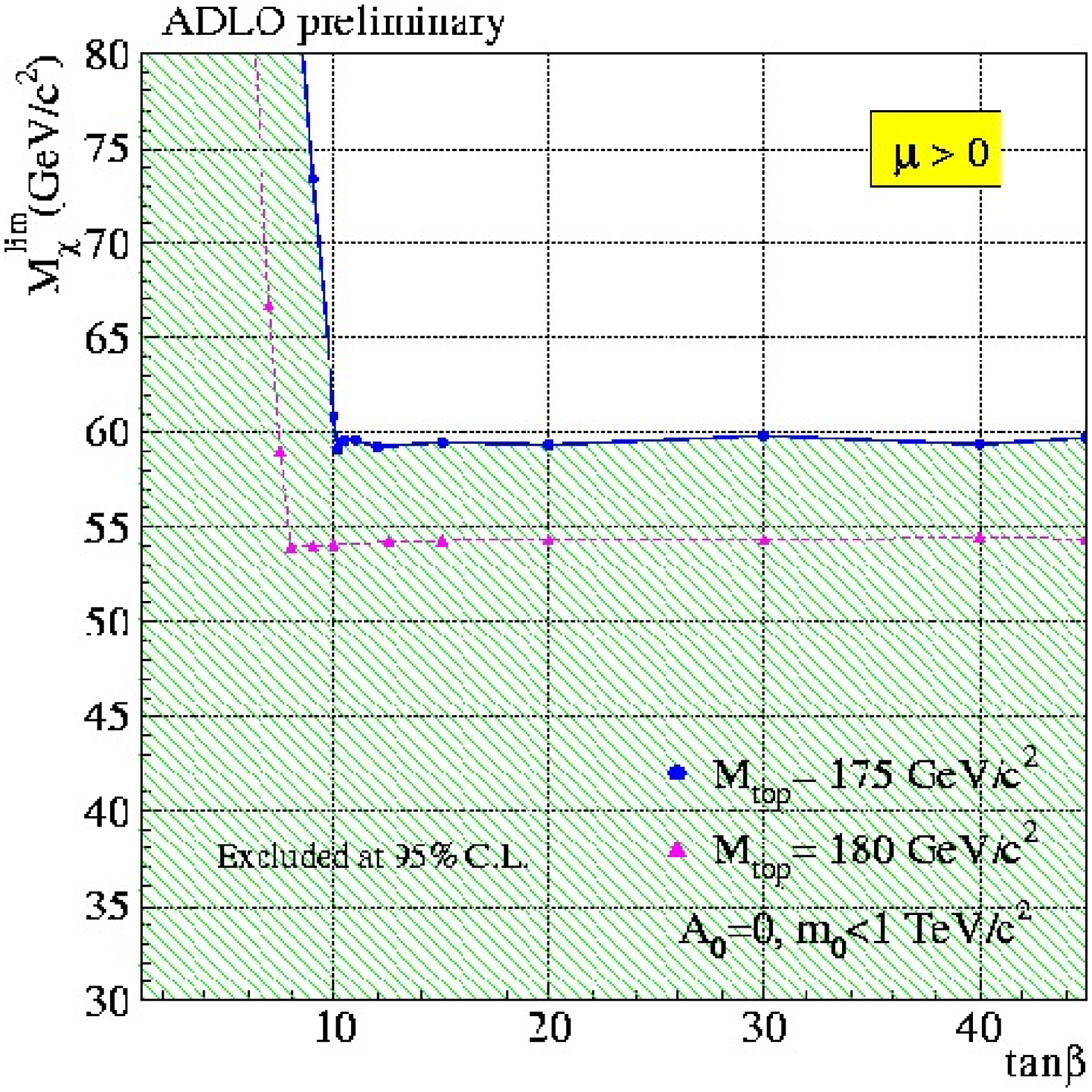}}&
\resizebox{80mm}{!}{\includegraphics{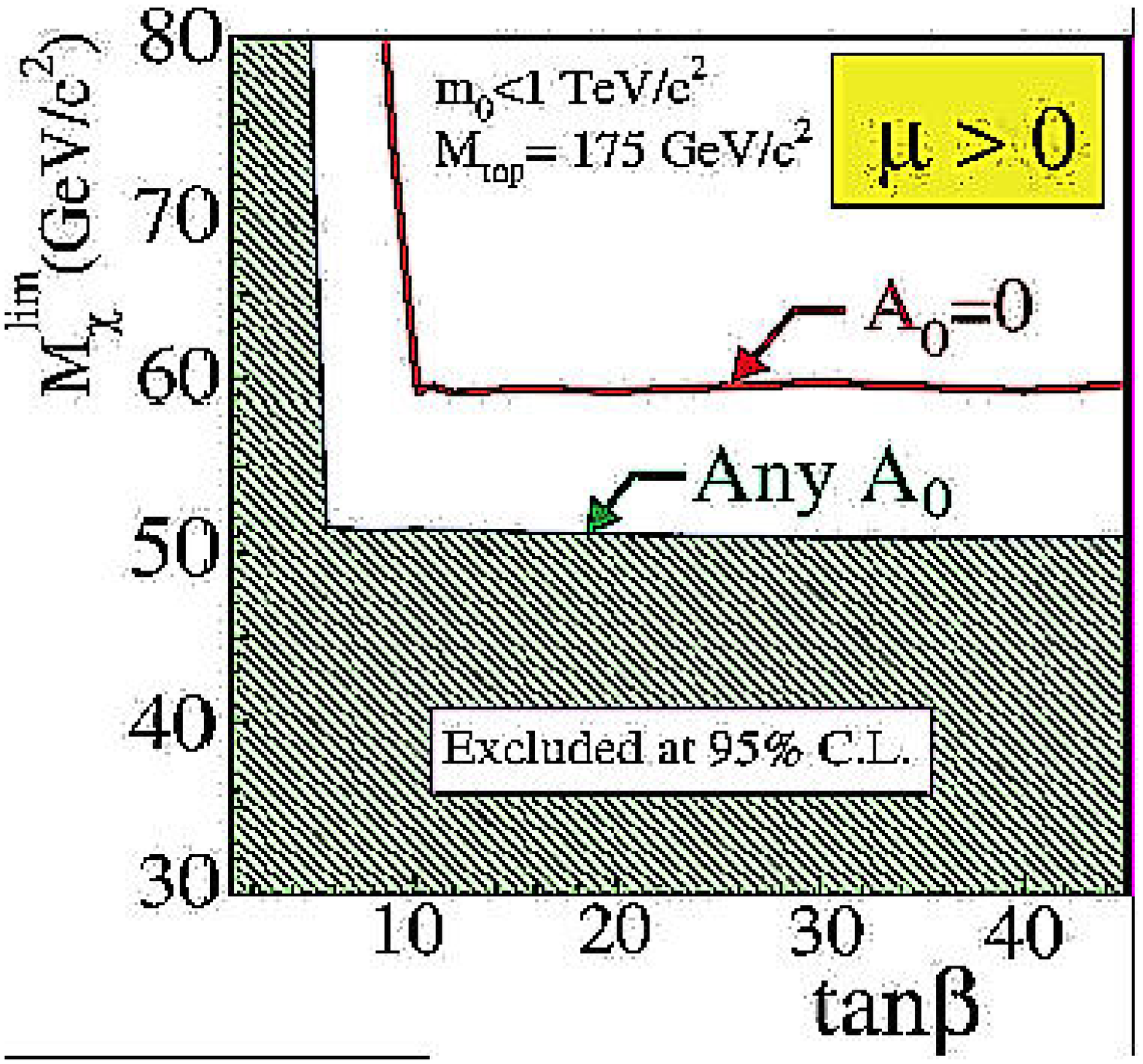}}\\
\end{tabular}
\caption[]{
Upper plots:
Exclusion regions in the mSUGRA scenario from Higgs and SUSY searches
at LEP for a range of $A_0$ values. On the right-hand side plot
effects of various searches are illustrated. Light shaded horizontal
region  is excluded by chargino searches,
hatched  bands are excluded by slepton searches (\selr\ and \stauo). 
Dark shading shows Higgs exclusion. Dedicated
neutralino search excludes area close to ``stau lsp'' region, complementing
the chargino search. Light shading shows the region where there is 
no good mSUGRA solutions (either due to charged LSP or no good
EWS breaking)
Left-hand side plot shows effect of changing $A_0$. For large negative
$A_0$ the region of the stau LSP grows, while the Higgs exclusion shrinks.
Lower plots:
The lower limit  on the mass of the lightest
neutralino, \XN{1}, in mSUGRA \protect{\cite{susywg}}.
Both plots are for positive $\mu$
which represents a more conservative scenario.
The  left-hand side plot illustrates the change of the limit
with the change of the top mass ($m_{top}=180.0$~\GeVcc,$m_{top}=175$ \GeVcc).
The right-hand side plot  shows the limit  obtained changing $A_0$ in
the bounds allowed by none of the third family sfermions
become tachyonic or the LSP.
The LSP limit degrades in this case down
to the one set by chargino searches and neutralino searches for $\tanb>15$.
}
\label{fig:msugraadlo}
\end{center}
\end{figure}

    Exclusion regions in the mSUGRA scenario obtained by the LEP
SUSY working group can be seen on figure \ref{fig:msugraadlo} for
an example value of \tanb\ and for a range of $A_0$ values.

Excluded regions in $m_{1/2}$ and $m_0$ can be translated  into limits on
$\MXN{1}$ , $\MXC{1}$ and other sparticles. Limits on $\MXN{1}$ obtained
by LEP SUSY WG \cite{susywg} are illustrated
on figure \ref{fig:msugraadlo} for several values of $A_0$
and $m_{top}$. $\MXC{1}$ is close to 2$\MXN{1}$.

As shown in 
\cite{mytevpaper} for large negative values of $A_0$  Higgs searches do
not exclude higher $m_{1/2}$ than the chargino searches already at moderate
\tanb\, thus the limit on the neutralino mass is set at the value
of around  50~\GeVcc\ by the
chargino searches, with neutralino and slepton searches covering
the stau and sneutrino hole.

\begin{figure}[!hbt]
\begin{center}
\resizebox{100mm}{!}{\includegraphics{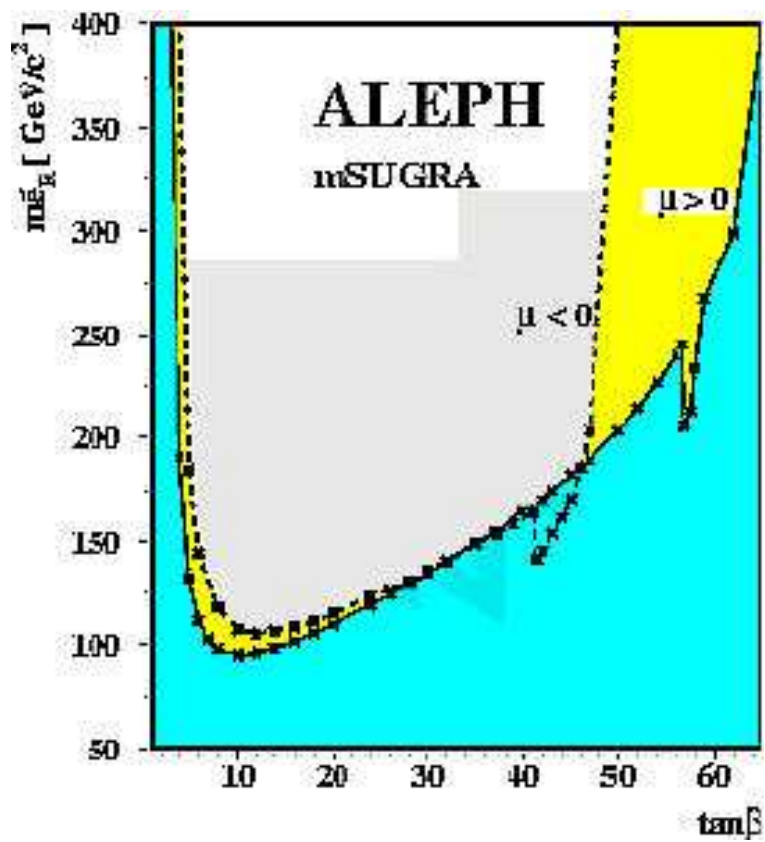}}
\caption[]{
The lower limit at 95~\% confidence level on the mass of the 
right-handed selectron, in mSUGRA \protect{\cite{susywg}} with
$A_0=0$. At low \tanb\ light selectrons are excluded by the limit
on the Higgs mass (which imposes a limit on $m_{1/2}$) 
and at large \tanb\ low $m_0, m_{1/2}$ values are exluded by stau
being the LSP. Both ``end'' of the selectron exlusion depend
on the $A_0$ value (see text).}
\label{fig:msugsel}
\end{center}
\end{figure}

ALEPH \cite{alephsel} obtained limits on selectron (\selr,\sell) and sneutrino
masses within mSUGRA for $A_0=0$. An example \mselr\ limit  as
a function of \tanb\ is shown on figure
\ref{fig:msugsel}. 
The limit is set by the Higgs searches at low \tanb. 
At high \tanb\, where the stau mixing is important also for lower $m_{1/2}$
the $m_0$ (and thus selectron mass) is pushed up by the requirement that
the stau is not the LSP. Both the Higgs exclusion and stau-LSP region
depend on the value of $A_0$.\\
   Similar limits for sleptons for several  values of $A_0$
are presented in \cite{mytevpaper}) along with limits on squarks and
gluino.

\section{Summary}
\label{sec:summary} 
LEP places relevant direct and indirect limits on the masses of nearly
all predicted sparticles. Direct limits are typically limited
by the kinematic reach of LEP and are valid for a specific decay channel
of a sparticle. Indirect limits often reach beyond the kinematic limit,
and are valid for all decays appearing in a specific more constrained
version of the MSSM. However, they make use of relations between the
sparticle masses, which are specific for the model in question (CMSSM or
mSUGRA).

\section{Acknowledgements}
\vskip 3 mm
I would like to thank the  Organizers of this enjoyable
Conference for the invitation.\\
 

\end{document}